# A High Frequency Trade Execution Model for Supervised Learning[*]

Matthew Dixon

Stuart School of Business
Illinois Institute of Technology
10 West 35th Street
Chicago, IL 60616

December 5, 2017

## Abstract

This paper introduces a high frequency trade execution model to evaluate the economic impact of supervised machine learners. Extending the concept of a confusion matrix, we present a 'trade information matrix' to attribute the expected profit and loss of the high frequency strategy under execution constraints, such as fill probabilities and position dependent trade rules, to correct and incorrect predictions. We apply the trade execution model and trade information matrix to Level II E-mini S&P 500 futures history and demonstrate an estimation approach for measuring the sensitivity of the P&L to the error of a Recurrent Neural Network. Our approach directly evaluates the performance sensitivity of a market making strategy to prediction error and augments traditional market simulation based testing.

## 1 Introduction

High frequency trading has been characterized as an arms race with 'Red Queen' characteristics (Farmer and Spyros, 2012). This perpetual state of needing to invest in infrastructure just to maintain a competitive advantage is an artifact of a 'flawed' continuous limit order book market design currently predominately deployed by financial exchanges (Budish et al., 2015). It is improbable, even impossible, that many market participants can sustain a competitive advantage through the sole reliance on low latency trade execution systems. But rather than looking to researchers, regulators and exchanges to overhaul the auctioning process, many high frequency trading firms have set out to leverage their technical might in other ways. The growth in volume of market data, advances in computer hardware and commensurate prominence of supervised learning in other disciplines, have spurred the exploration of supervised learning for price discovery.

Modern financial markets facilitate the electronic trading of financial instruments through an instantaneous double auction. At each point in time, the market demand and the supply can be represented by an electronic limit order book, a cross section of orders to execute at various price levels away from the market price. The price levels closest to the market price define the 'inside market' and is the most actively traded. The market price is closely linked to its liquidity - that is the immediacy in which the instrument can be converted into cash. The liquidity of markets are characterized by their depth, the total quantity of quoted buy and sell orders about the market price and is hence constantly evolving in response to trading activity (Bloomfield et al., 2005).

A participant enters into a trade by submitting an order to a queue and either waits up to a few seconds for the order to be filled or cancels the order. This type of trading adds liquidity and is said to be 'making a market'. Market making is a primary function of proprietary traders that are making markets algorithmically and from buy-side institutions that are submitting limit orders as part of 'slice and dice' algorithms (Hendershott et al., 2011). A participant willing to pay a premium to trade at the best price can bypass the queue and is said to be 'market taking'.

---

[*]An early version of this paper was presented at the Machine Learning Mini-Symposium of the 2016 SIAM Conference on Financial Mathematics and Engineering, Austin, TX. The author would like to thank Prof. Brian Peterson, University of Washington and Partner at DV Trading, for many helpful comments.

Liquid markets are attractive to market participants as they permit the near instantaneous execution of large volume trades at the best available price, with marginal price impact. However, sometimes a large market order, or a succession of smaller markets orders, will consume an entire price level. This is why the market price fluctuates in liquid markets - an effect often referred to by practitioners as a 'price-flip'. Price level consumption is followed by an initial widening of the bid-ask spread which quickly reverts as market makers exploit it, leading to a new mid-price.

Microstructure researchers seek to evaluate how the increased information stored in the limit order book informs price discovery and ultimately translates to consistent economic utility from trading. There appears to be no consensus on the extent to which limit order books convey predictive information. Early seminal papers studying equities, including Glosten (1994); Seppi (1997), state that the limit orders beyond the insider market contain little information. In contrast, several other studies state that such limit orders are informative (Parlour, 1998; Bloomfield et al., 2005; Cao et al., 2009; Zheng et al., 2013; Kearns and Nevmyvaka, 2013; Cont et al., 2014; Palguna and Pollak, 2014). In particular, Cao et al. (2009) study the information content of a limit-order book from the Australian Stock Exchange. They found that the book's contribution to price discovery is approximately 22% while the remaining comes from the first level data and transaction prices. They also demonstrate that order imbalances between the demand and supply exhibit a statistically significantly relationship to short-term future returns. There is growing evidence that the study of microstructure is critical to finding longer term relations and even cross-market effects (Dobrislav and Schaumburg, 2016).

Many quantities, such as the probability of price movements given the state of the limit order, are relevant for trading and intraday risk management. The complex relation between order book dynamics and price movements has been the focus of econometric and stochastic modeling (see Engle and Russell (1998); Cont et al. (2010b, 2014); Cont and de Larrard (2013); Chavez-Casillas and Figueroa-Lopez (2017) and references therein). For analytic tractability, these models assume a data generation process and typically estimate quantities based on asymptotic limits of diffusion processes.

In a Markovian setting, and under further modeling assumptions, such as the treatment of the arrival rate of market orders as a poisson process, homogenous order sizes, and the assumption of independence of cancellations and orders, a probability of an up-tick is derived. However, these modeling assumptions made are likely too strong for describing micro-scale book dynamics (sub 1ms). At this scale, price is not Markovian, increments are neither independent nor stationary and depend on the state of the order book. Attempts to relax the Markovian assumption, using for example the 'heavy traffic' approximation approach (Cont and de Larrard, 2010; Chavez-Casillas and Figueroa-Lopez, 2017) are best suited for meso-scale analysis of the price movements, but not micro-scale.

Guided by the reduced order book models of Cont and de Larrard (2013), our approach selects similar exogenous variables. In particular, we treat queue sizes at each price level as the independent variables. We additionally include properties of market orders, albeit in a form which we have observed to be most relevant to prediction the direction of price movements. In sharp contrast to stochastic modeling, we do not impose conditional distributional assumptions on the independent variables (a.k.a. features) nor assume that price movements are Markovian.

Most of the aforementioned predictive modeling studies rely on regression for explanatory power of liquidity on the *continuous* volume weighted average price (VWAP), a.k.a. 'smart price'. The utility of the smart price is limited for high frequency trading. So called 'market makers' quote limit orders on both sides of the market in attempt to capture the spread. Their inability to pre-empt a price flip, by adjusting their quotes, typically results in adverse price selection and a loss of profit. A change in the smart price does not imply a price-flip. For example, a change in the volume of the best bid quantity will result in a change in the smart price, but not necessarily a change in the mid-price, the latter effect is attributed to price level consumption from incoming market orders. The successful prediction of a price flip, and not the change in smart price, can therefore be directly used to avoid adverse price selection. Sudden price flips in the the tick data are hard to capture with traditional modeling techniques which rely on instantaneous inside market liquidity imbalance alone.

Breiman (2001) describes the two cultures of statistical modeling when deriving conclusions from data. One assumes a data generating process, the latter uses algorithmic models, treating the data mechanism as unknown. Machine learning falls into the algorithmic class of reduced model estimation procedures. It is designed to provide predictors in complex settings where relations between input and output variables are nonlinear and input space is often high dimensional. A number of researchers have applied machine learning methods to the study of limit order book dynamics (Kearns and Nevmyvaka, 2013; Kercheval and Zhang, 2015; Sirignano, 2016; Dixon et al., 2017; Dixon, 2017).

This paper takes an algorithmic approach to predicting the next event price-flip from a short sequence of observations of limit order book depths and market orders. We choose a spatio-temporal representation (Sirignano, 2016; Dixon, 2017) of the limit order book combined with history of the market orders as the predictors. Our approach solves a sequence classification problem - a short sequence of observations of book depths and market orders can be classified into directional mid-price movement. A sequence classifier offers potential significant benefit to market participants. For example, a market maker can use the classifier to continuously adjust the quotes, potentially reducing the likelihood of adverse price selection. Sequence classification has been considered elsewhere in the literature for lower frequency price movement prediction from historical prices (Leung et al., 2000; Dixon et al., 2016). The novelty of our approach therefore arises from the application of a recurrent neural network classifier to a spatio-temporal representation of the limit order book combined with market order history in order to predict price-flips.

Training a recurrent neural network architecture can be performed with stochastic gradient descent (SGD) which learns the weights and offsets in an architecture between the layers. Drop-out (DO) performs variable selection (Srivastava et al., 2014). RNNs rely on a moderate amount of training time series data together with a flexible architecture to 'match' in and out of sample performance as measured by mean error, area under the curve (AUC) or the F1 score, which is the harmonic mean of precision and recall.

**Economic value** The aforementioned theoretical and empirical research articles partially address the question of whether limit orders contain information beyond the best bid and ask prices. Through the proliferation of electronically traded exchanges, traders can use large numbers of variables, often available at every tick, when making trading decisions. Researchers are also able to use techniques that are more sophisticated than the standard time series analysis to forecast future price movements. However, until recently, there have been few studies focusing on whether this information can be efficiently and systematically translated to consistent economic profits. Despite finding statistically significant explanatory variables describing the structure of the limit order book, Kozhan and Salmon (2012) and Kearns and Nevmyvaka (2013) conclude, in their respective studies of the FX and Equity markets, that the information content of the limit order book does not seem to translate to greater economic profits through different high frequency trading rules. More precisely, these authors arrive at a similar conclusion that limit order book data alone is not robust enough to justify market taking.

**Confusion matrices** There are several techniques for measuring the performance of classification based supervised learning models including the ROC curve, the confusion matrix, the F-score (see for example Bishop (2006) or Hastie et al. (2001)). The confusion matrix remains one of the most widest techniques. Each column of the matrix represents the instances in a predicted class while each row represents the instances in an actual class. It is often instructive to monitor the true positive and true negative rates, which are ratios of true positive instances of a predicted positive class to the total instances of actual positives and true negative instances of a predicted negative class to the total instances of actual negatives respectively. Oftentimes, in application, the significance of the true positive rate may not be the same as that of the true negative rate. For example, in fraud detection, there may be a higher penalty associated with falsely predicting a negative class when the instance is in fact positive, than predicting a positive when the instance is in fact negative. This construct is analogous to Type I and Type II errors in the statistics literature. It is common practice in the decision sciences to oftentimes weight the true positive and negative rates by their economic significance accordingly. Borrowing such a concept, made concrete within the context of fraud detection in Bhowmik (2008), we introduce a trade information matrix to characterize the economic impact of false positives and false negatives.

## 1.1 Overview

The main contribution of this paper is to present a 'trade information matrix' to attribute the expected profit and loss of tick level predictive classifiers under execution constraints, such as fill probabilities and position dependent trade rules, to correct and incorrect predictions. We introduce and apply a trade execution model for evaluating market making strategies and use Level II E-mini S&P 500 futures history to estimate the terms in the trade information matrix. Using exchange matching engine rules, this trade execution model estimates the queue position of a reference order and determines whether it is filled on arrival of new market orders and limit order cancellations. The probability of a fill is estimated based on its size, side, level and time that it was placed. Such probabilistic measures of liquidity constraints govern how the quotes placed by a strategy generate expected P&L and are conveniently expressed in a trade information matrix. This trade information matrix is then simply multiplied by the confusion matrix of the classifier to assess the economic impact of Type I and II error.

We begin in the next section by introducing supervised machine learning and our preferred classification approach, recurrent neural networks. Section 3 motivates the application of supervised learning to market making and then presents a trade execution model. Section 4 presents the trade information matrix for measuring strategy performance under error. Section 4.2 describes the preparation of the Level II data used to train the classifier, referred to as the 'feature set'. Section 5 presents results measuring the performance of the classifier and demonstrates the estimation of the trade information matrix to market data. Our results show the degree of error tolerance in a market making strategy that uses the prediction signal. The results also compare the strategy with a 'blind' strategy that does not use prediction and describes the factors when the former is more favorable. For completeness, Section 5.2 presents further results demonstrating that there is little gain from re-training the model on a frequent basis; (ii) that there are distinct intra-day classifier performance trends; and (iii) classifier accuracy quickly erodes with the length of prediction horizon. Finally Section 6 concludes with comments and further research questions aimed at addressing the practicality of using supervised learning and the trade information matrix.

## 2 Supervised Learning

Supervised learning addresses a fundamental prediction problem: Estimate a nonlinear predictor, $\hat{Y}(X)$, of an output, $Y$, given a high dimensional input matrix $X = (X^{(1)}, \ldots, X^{(P)})$ of $P$ variables. Supervised learning can be simply viewed as the study and estimation of an input-output map of the form

$$Y = F(X) \text{ where } X = (X^{(1)}, \ldots, X^{(P)}).$$

The output variable, $Y$, can be continuous, discrete or mixed. For example, in a classification problem, $F : X \to Y$ where $Y \in \{1, \ldots, K\}$ and $K$ is the number of categories. When $Y$ is a continuous vector and $f$ is a semi-affine function, then we recover the linear model

$$Y = AX + b.$$

### 2.1 Sequence Learning

If the input-output pairs $\mathcal{D} = \{X_t, Y_t\}_{t=1}^{N}$ are auto-correlated observations of $X$ and $Y$ at times $t = 1, \ldots, N$, then the fundamental prediction problem can be expressed as a sequence prediction problem: estimate a nonlinear times series predictor, $\hat{Y}(\mathbb{X})$, of an output, $Y$, using a high dimensional input matrix of $T$ length sub-sequences $\mathbb{X}$:

$$y = F(\mathbb{X}) \text{ where } \mathbb{X}_t = seq_T(X_t) = (X_{t-T+1}, \ldots, X_t)$$

where $X_{t-j}$ is a $j^{th}$ lagged observation of $X_t$, $X_{t-j} = L^j[X_j]$, for $j = 0, \ldots, T-1$. Sequence learning, then, is just a composition of a non-linear map and a vectorization of the lagged input variables. If the data is i.i.d., then no sequence is needed (i.e. $T = 1$), and we recover the standard prediction problem.

## 2.2 Recurrent Neural Networks (RNNs)

RNNs are sequence learners which have achieved much success in applications such as natural language understanding, language generation, video processing, and many other tasks Graves (2013). We will concentrate on simple RNN models for brevity of notation.

A simple RNN is formed by a repeated application of a function $F_h$ to the input sequence $\mathbb{X}_t = (X_1, \ldots, X_T)$. For each time step $t = 1, \ldots, T$, the function generates a hidden state $h_t$ from the current input $X_t$ and the previous output $h_{t-1}$:

$$h_t = F_h(X_t, h_{t_1}) = \sigma(W_h X_t + U_h h_{t_1} + b_h), \tag{1}$$

for some non-linear activation function $\sigma(x)$.

When the output is continuous, the model output from the final hidden state, $Y = F_y(h_T)$, is given by the semi-affine function:

$$Y = F_y(h_T) = W_y h_T + b_y, \tag{2}$$

and when the output is categorical, the output is given by

$$Y = F_y(h_T) = \text{softmax}(F_y(h_T)), \tag{3}$$

where Y has a 'one-hot' encoding - a K-vector of zeros with 1 at a single position. Here $W = (W_h, U_h, W_y)$ and $b = (b_h, b_y)$ are weight matrices and offsets respectively. $W_h \in \mathbb{R}^{H \times P}$ denotes the weights of non-recurrent connections between the input $X_t$ and the $H$ hidden units. The weights of the recurrence connections between the hidden units is denoted by the recurrent weight matrix $U_h \in \mathbb{R}^{H \times H}$. Without such a matrix, the architecture is simply an unfolded single layer feed-forward network without memory and each observation $X_t$ is treated as an independent observation.

$W_y$ denotes the weights tied to the output of the hidden units at the last time step, $h_t$, and the output layer. If the output variable is a continuous vector, $Y \in \mathbb{R}^M$ then $W_y \in \mathbb{R}^{M \times H}$. If the output is categorical, with K states, then $W_y \in \mathbb{R}^{K \times H}$.

## 2.3 Training, Validation and Testing

To estimate and evaluate a learning machine, we start by controlled splitting of the data into training, validation and test sets. The training data consists of input-output pairs $\mathcal{D} = \{Y_t, X_t\}_{t=1-(T-1)}^N$. We then sequence the data to give $\mathcal{D}_{\text{seq}} = \{Y_t, \mathbb{X}_t\}_{t=1}^N$.

The goal is to find the machine sequence learner $Y = F(\mathbb{X})$, where we have a loss function $\mathcal{L}(Y, \hat{Y})$ for a predictor, $\hat{Y}$, of the output signal, $Y$. In many cases, there's an underlying probability model, $p(Y \mid \hat{Y})$, then the loss function is the negative log probability $\mathcal{L}(Y, \hat{Y}) = -\log p(Y \mid \hat{Y})$. For example, under a Gaussian model $\mathcal{L}(Y, \hat{Y}) = ||Y - \hat{Y}||^2$ is a $L^2$ norm, for binary classification, $\mathcal{L}(Y, \hat{Y}) = -Y \log \hat{Y}$ is the negative cross-entropy.

In its simplest form, we then solve an optimization problem

$$\underset{W,b}{\text{minimize}}\ f(W,b) + \lambda \phi(W, b)$$

$$f(W, b) = \frac{1}{N} \sum_{t=1}^{N} \mathcal{L}(Y_t, \hat{Y}(\mathbb{X}_t))$$

with a regularization penalty, $\phi(W, b)$.

Here $\lambda$ is a global regularization parameter which we tune using the out-of-sample predictive mean-squared error (MSE) of the model on the verification data. The regularization penalty, $\phi(W, b)$, introduces a bias-variance tradeoff. $\nabla \mathcal{L}$ is given in closed form by a chain rule and, through back-propagation on the unfolded network, the weight matrices $\hat{W}$ are fitted with stochastic gradient descent. See Rojas (1996); Graves (2013) for a further description of stochastic gradient descent as it pertains to recurrent neural networks.

# 3 High Frequency Trading

We shall begin by informally motivating why machine learning is useful for high frequency trading. Consider the following scenario where a market maker offers to sell at price $x_t + s/2$ and buy at price $x_t - s/2$ in an attempt to capture the spread $s$, where $x_t$ is the mid-price at time $t$. The market maker runs the risk of an adverse mid-price movement between the fill of the sell order and the buy order. For example, a buy market order matching with the entire resting ask quantity, at price $x_t + s/2$, effectively results in an up-tick of the mid-price, $x_{t+1} = x_t + s$. The market maker has sold at $x_t + s/2$ but can very likely not buy back at $x_t - s/2$ as the bid is no longer at the inside market. Instead the market marker may systematically be forced to cancel the bid and buy back at a price higher than $x_t - s/2$, taking a loss.

Figure 1 (left) illustrates a typical mechanism resulting in an adverse price movement. A snapshot of the limit order book at time $t$, before the arrival of a market order, and after at time $t+1$ are shown in the left and right panels respectively. The resting orders placed by the market marker are denoted with the '+' symbol- red denotes a buy limit order and blue denotes a sell limit order. A buy market order subsequently arrives and matches the entire resting quantity of best ask quotes. Then at event time $t+1$ the limit order book is updated - the market maker's sell limit order has been filled (blue minus symbol) and the buy order now rests away from the inside market. The mechanism is analogous for a down-tick of the book.

In the above illustrated example, the market maker is assumed to have no knowledge of the future and therefore does not pre-empt the price movement by adjusting their quotes accordingly. But suppose now that the market maker reasonably suspects an up-tick and has the capacity to avoid a loss resulting from the adverse price movement.

Figure 1 (right) illustrates the result of a market maker acting to prevent losing the spread -the sell limit order is adjusted to a higher ask price. In this illustration, the buy limit order is not replaced and the market maker may capture a tick more than the spread. However, it's unlikely that the buy limit order will be filled and hence may also need to be adjusted up to the inside market, albeit with loss of queue position. Such trade-offs are the focus of execution optimization and, although important, are not considered further here.

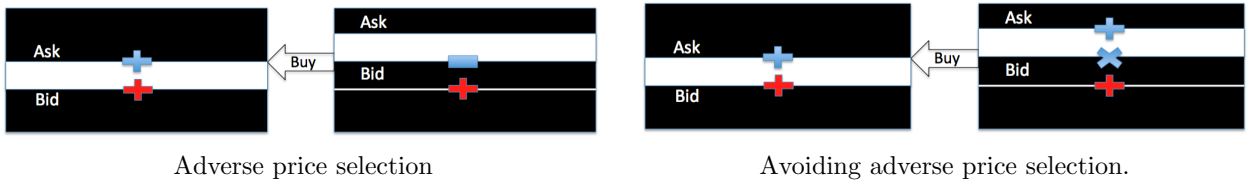

Adverse price selection.  Avoiding adverse price selection.

Figure 1: *(left) Illustration of a typical mechanism resulting in adverse price selection. A snapshot of the limit order book is taken at time $t$. Limit orders placed by the market marker are denoted with the '+' symbol- red denotes a buy limit order and blue denotes a sell limit order. A buy market order subsequently arrives and matches the entire resting quantity of best ask quotes. Then at event time $t+1$ the limit order book is updated. The market maker's sell limit order has been filled (blue minus symbol) and the buy order now rests away from the inside market. (Right) A pre-emptive strategy for avoiding adverse price selection is illustrated. The sell limit order is requoted at a higher ask price. In this case, the buy limit order is not replaced and the market maker may capture a tick more than the spread if both orders are filled.*

## 3.1 Level-2 representation of a limit order book

As mentioned in the previous section, the amount of liquidity in the market can be characterized by the cross-section of book depths and this liquidity evolves in response to trading activity. Note that the purpose of this paper is not to explicitly model the dynamics of the limit order book as in, for example, Cont et al. (2010a,b). Instead, we seek to build a simple exchange simulator sufficient to measure the performance of a trading strategy as a function of model prediction error. More formally, we represent the state of the limit order book by

- the n-vector of bid prices $\mathbf{s}^b := (s_t^{b,1}, \ldots, s_t^{b,n})$ and the ask prices $\mathbf{s}^a := (s_t^{a,1}, \ldots, s_t^{a,n})$ at all levels of the order book. We use the convention that the $i = 1$ component represents the best bid and ask and the $i > 1$ represents each level away from the inside market;

- the n-vector of bid queues $\mathbf{q}_t^b := (q_t^{b,1}, \ldots, q_t^{b,n})$, where $q_t^{b,i}$ denotes the depth of each bid level $i$ at time $t$; and

- the n-vector of ask queues $\mathbf{q}_t^a := (q_t^{a,1}, \ldots, q_t^{a,n})$ where $q_t^{a,i}$ denotes the depth of each bid level $i$ at time $t$.

The state of the limit order book is thus described by $\mathcal{X}_t := (\mathbf{s}_t^b, \mathbf{s}_t^a, \mathbf{q}_t^b, \mathbf{q}_t^a)$ which takes values in the discrete state space $\delta \cdot \mathbb{Z}^{2n} \times \mathbb{N}^{2n}$ for some tick size $\delta << 1$. Note, for avoidance of doubt, that this notation still permits the bid-ask spread to change - this occurs when one or more of the price levels at the inside market has zero depth.

The state $\mathcal{X}_t$ of the order book is modified by the following order book events arriving at time $t$:

- limit orders (at the bid or ask) of size $\mathbf{L}_t^b := (L_t^{b,1}, \ldots, L_t^{b,n})$ and $\mathbf{L}_t^a := (L_t^{a,1}, \ldots, L_t^{a,n})$;

- market orders (to buy or sell) of size $M_t^b$ and $M_t^s$. These are referred to by practitioners as 'aggressors'; and

- cancelations of limit orders of size $\mathbf{C}_t^b := (C_t^{b,1}, \ldots, C_t^{b,n})$ and $\mathbf{C}_t^a := (C_t^{a,1}, \ldots, C_t^{a,n})$.

When the bid or ask queue is depleted, the price moves up or down to the next level of the order book. More precisely,

1. When the best bid queue is depleted, the mid-price decreases by half a tick. The bid-ask spread is temporarily one and a half ticks;

2. Quotes subsequently arrive at a new lower ask level so that the bid-ask spread returns to a tick - the mid-price has now decreased by another half-a-tick;

3. This full tick change in the mid-price, inclusive of the intermediate state, shall be referred to as a 'price-down-flip' or just 'price-flip'. Practitioner's also use the terminology 'down-tick' or the price is said to 'tick-down'.

The converse mechanism, when the best ask queue is depleted, is referred to a 'price-up-flip'. The following example illustrates a price-down-flip in response to the arrival of a sell aggressor.

**Example: Order book update** Figure 2 illustrates a typical mechanism resulting in mid-price movement. The charts on the left and right respectively show the limit order book of ESU6 before and after the arrival of a large sell aggressor. The aggressor is sufficiently large to match all of the best bids. Once matched, the limit order is updated with a lower best bid of \$2175.5. The gap between the best ask and best bid would widen if it weren't for the arrival of 23 new contracts offered at a lower ask price of \$2175.75. The net effect is a full down-tick of the mid-price.

Using the notation introduced in this section, Table 1 records the limit order book before and after the arrival of the sell aggressor.

| time  | $\mathcal{X}_t^1$            | $M_t^s$ | $L_t^{a,1}$ |
|-------|------------------------------|---------|-------------|
| $t_0$ | $(2175.75, 2176.0, 103, 82)$ | 0       | 0           |
| $t_1$ | $(2175.5, 2176.0, 177, 82)$  | 103     | 0           |
| $t_2$ | $(2175.5, 2175.75, 177, 23)$ | 0       | 23          |

Table 1: *This table shows the limit order book of ESU6 before and after the arrival of the sell aggressor, as illustrated in Figure 2.*

The focus of the remainder of the paper is to develop and demonstrate a practical approach for measuring the economic impact of the ability to predict these price-flips. We use supervised learning to predict the price movements and then combine a simple model for estimating the queue position, fill execution, and for a given market making strategy, estimate the expected P&L as a function of prediction error. The market making strategy exploits predicted price movements.

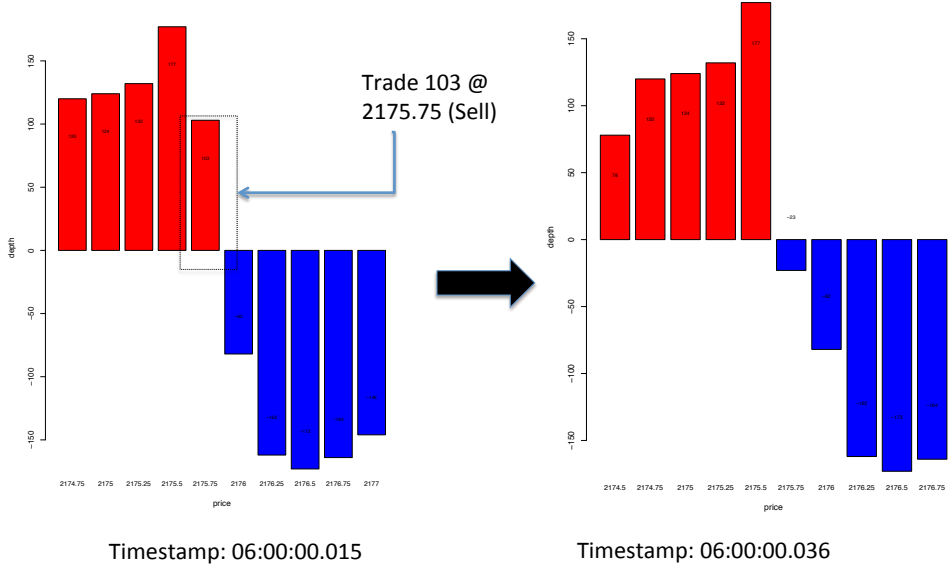

Figure 2: *This figure illustrates a typical mechanism resulting in mid-price movement of ESU6. The charts on the left and right respectively show the limit order book before and after the arrival of a large sell aggressor.*

## 3.2 Trade Execution

Given a description of the order book and its updates, we seek to evaluate whether a quote placed by a market participant is filled over some time horizon. The ability to accurately model this action from this data is central to evaluating the performance of any high frequency trading strategy and, specifically for this paper, the economic impact of a supervised learning based trade decision.

Consider a market participant that places a limit buy order at the best bid price. This order, hereon be referred to as the 'reference order', is received by the exchange at time $t_0$. Suppose that the matching engine receives $n$ market or limit orders which affect the queue position of the best bid level in the limit order book over the subsequent interval $\tau := t - t_0$. These orders are either sell market orders, buy limit order cancellations or new buy limit orders. The set of arrival times of all orders, including the reference order, is $\mathbf{t} := \{t_0, \ldots, t_n\}$. Implicit in our notation, is the constraint that no two orders can arrive and be processed by the matching engine at exactly the same time.

Denote the sets of order sizes as

- sell market orders: $M_\tau^s := \{M_t^s \mid t \in \mathbf{t}^s\}$ which arrive at times $\mathbf{t}^s \subseteq \mathbf{t}$;
- buy market orders: $M_\tau^b := \{M_t^b \mid t \in \mathbf{t}^b\}$ which arrive at times $\mathbf{t}^b \subseteq \mathbf{t}$;
- bid order cancellations: $C_\tau^{b,i} := \{C_t^{b,i} \mid t \in \mathbf{t}^{c,i}\}$;
- ask order cancellations: $C_\tau^{a,i} := \{C_t^{a,i} \mid t \in \mathbf{t}^{c,i}\}$;
- bid orders: $L_\tau^{b,i} := \{L_t^{b,i} \mid t \in \mathbf{t}^{b,i}\}$; and
- ask orders: $L_\tau^{a,i} := \{L_t^{a,i} \mid t \in \mathbf{t}^{a,i}\}$.

Further denote the queue position at time $t$ of a limit order arriving at time $s$ as $\phi(t, s)$. With slight abuse of notation, denote $\mathcal{X}_0^j$ as the state of the first $j$ levels of the order book at time $t_0^-$. The subset of events arriving in period $\tau$ that change $\mathcal{X}_0^j$, are denoted by

$$\mathcal{D}_\tau^{b,j} = (M_\tau^s, \{C_\tau^{b,i}\}_{i=1}^j, \{L_\tau^{b,i}\}_{i=1}^j) \qquad \mathcal{D}_\tau^{a,j} = (M_\tau^b, \{C_\tau^{a,i}\}_{i=1}^j, \{L_\tau^{a,i}\}_{i=1}^j) \qquad (4)$$

and the union of these sets represents all book updates $\mathcal{D}_\tau^j = \mathcal{D}_\tau^{a,j} \bigcup \mathcal{D}_\tau^{b,j}$. Clearly, book updates at the inside market are a subset of book updates resolved to outer levels:

$$\mathcal{D}_\tau^1 \subset \mathcal{D}_\tau^2 \subset \cdots \subset \mathcal{D}_\tau \equiv \mathcal{D}_\tau^n. \qquad (5)$$

Similarly the state of the top level of the book is a subset of the state including the outer levels,

$$\mathcal{X}_t^1 \subset \mathcal{X}_t^2 \subset \cdots \subset \mathcal{X}_t \equiv \mathcal{X}_t^n.$$

It is instructive to illustrate the notation with an example which is closely related to the one illustrated in Figure 2.

**Example** Suppose that prior to time $t_0$ the top of the book is $\mathcal{X}_0^1 = (2175.75, 2176.0, 102, 82)$. The following sequence of events occur:

1. At time $t_0$, a one lot best bid order, $L_{t_0}^{b,1} = 1$, is received by the exchange.

2. As in the previous example, a sell market order of size $M_{t_1}^s = 103$ is received by the exchange at time $t_1$ and clears the best buy level. The best bid price and quantity is updated.

3. Then at time $t_2$, a limit order of size 23 is quoted at a new best ask level, restoring the spread to a tick (0.25).

The books updates, $\mathcal{D}_\tau^1$, and the current state of the book, $\mathcal{X}_t^1$, are enumerated in Table 3.2.

| time | $\mathcal{X}_0^1$ | $M_\tau^s$ | $M_\tau^b$ | $C_\tau^{b,1}$ | $C_\tau^{a,1}$ | $L_\tau^{b,1}$ | $L_\tau^{a,1}$ | $\mathcal{X}_t^1$ |
|---|---|---|---|---|---|---|---|---|
| | | \multicolumn{6}{c|}{$\Omega_\tau^1$} | |
| $t_0^-$ | (2175.75, 2176.0, 102, 82) | {} | {} | {} | {} | {} | {} | (2175.75, 2176.0, 102, 82) |
| $t_0$ | (2175.75, 2176.0, 102, 82) | {} | {} | {} | {} | {1} | {} | (2175.75, 2176.0, 103, 82) |
| $t_1$ | (2175.75, 2176.0, 102, 82) | {103} | {} | {} | {} | {1} | {} | (2175.5, 2176.0, 177, 82) |
| $t_2$ | (2175.75, 2176.0, 102, 82) | {103} | {} | {} | {} | {1} | {23} | (2175.5, 2175.75, 177, 23) |

Table 2: *This table shows how the state of the top of the book $\mathcal{X}_t^1$ is updated by data $\mathcal{D}_\tau^1$.*

**Remark 3.2.1** (market data). *The evolution of the state of the book shown in Table 3.2 assumes rules for pairing the market order with the limit order. Therefore, in order to simulate the state evolution, these rules must be prescribed but are, in fact, complex and based on the type of market data available and the exchange. Moreover, in many cases, the market data is not sufficiently complete to specify the evolution of the state and a model must be introduced to estimate the probability of order execution. Without such a model, the economic impact of a decision to trade can not be known and is therefore central to algorithmic trading.*

### 3.3 Execution Model

Informally, we assume that there exists a state update function

$$H_\mathcal{M} : \delta \cdot \mathbb{Z}^{2n} \times \mathbb{N}^{2n} \to \delta \cdot \mathbb{Z}^{2n} \times \mathbb{N}^{2n} \qquad (6)$$

that uses new data to evolve the state

$$\mathcal{X}_t = H_\mathcal{M}(\mathcal{X}_0; \mathcal{D}_\tau). \qquad (7)$$

The function depends on heuristics for updating the state given new data, $\mathcal{D}_\tau$. These heuristics depend on the matching rule, $\mathcal{M}$- the exchange defined procedure for finding pairs or groups of orders that are executed against each other. The most common types of matching are FIFO and pro-rata, each leading to different scenarios under which a limit order may gain queue position.

We shall assume that queue position of a limit order place at time $s$ is given by a function $\phi_s : \mathbb{R}^+ \times \mathbb{Z} \to \mathbb{N}$, so that $\phi_{t,s} := \phi_t(L_s; \mathcal{X}_{s-}, \mathcal{D}_\tau)$ gives its queue position at time $t \geq s$. These matching rules are different for each exchange and are occasionally updated.

The reference bid (or ask) order $L_0$ received by the exchange at time $t_0$ is filled (a.k.a. executed) by time $t_n$ if both criteria are met:

1. All bids (respectively offers) at the best level in the limit order book with higher queue priority are either cancelled or all these higher priority limit orders, including the reference order, are matched with one or more sell (respectively buy) aggressors arriving over the interval $\tau$. Our order is partially filled if, under the same set of events, the exchange only matches a portion of the reference order.

2. Any reference order cancellation is not received at or before time $t_n$.

We quantify the extent to which $L_0$ is filled given the triple $(L_0, \mathcal{X}_0, \mathcal{D}_\tau)$. Since the queue position of $L_0$ in unknown, we introduce a parameterization representing the degree of conservativeness in order execution. We introduce the trade-to-book ratio as a measure of the trade size relative to resting quantity in front of and including an order placed at time $t_0$. This ratio is scaled so that a value of at least unity indicates a complete fill. Otherwise, the ratio indicates that the current trade size is inadequate to completely fill the order. It is further possible to determine the lower bound corresponding to a partial fill of the reference order.

**Definition 3.3.1** (Trade-to-Book Ratio). *In the event of a sell market order arriving at time $t$, the trade-to-book ratio of a level $j$ bid limit order, $L_0^{b,j}$, placed at time $t_0$ is a function $\mathcal{R} : \mathbb{R}^+ \times \mathbb{N} \to \mathbb{R}^+$ of the form*

$$\mathcal{R}_t(L_0^{b,j}; \mathcal{D}_\tau^{b,j}, \omega) = \frac{M_t^s}{Q_0^{b,j} - \left(\sum_{u \in \mathbf{t}^s} M_u^s + \omega \sum_{i=1}^j \sum_{u \in \mathbf{t}^{c,i}} C_u^{b,i} - \sum_{i=1}^j \sum_{t \in \mathbf{t}^{b,i}} \mathbb{1}_{\{\phi_{u,u} < \phi_{u,t_0}\}} L_u^{b,i}\right)} \quad (8)$$

where

- $Q_0^{b,j} := \sum_{i=1}^j q_{t_0}^{b,i}$ *denotes the sum of the depths of the queue at time $t_0$ up to the $j^{th}$ bid level;*

- $\sum_{u \in \mathbf{t}^s} M_u^s$ *are subsequent set of sell market orders arriving at times $\mathbf{t}_s$;*

- $\sum_{u \in \mathbf{t}^{c,i}} C_u^{b,i}$ *is the subsequent of level $i$ bid orders cancelled at times $\mathbf{t}^{c,i}$;*

- $\mathbb{1}_{\{\phi_{u,u} < \phi_{u,t_0}\}}$ *is an indicator function returning unity if a subsequent limit order placed at time $u$ has higher queue priority than the time $t_0$ reference limit order; and*

- $\omega \in [0,1]$ *is an unknown cancellation parameter which denotes the proportion of cancellations of orders with higher queue priority than the reference limit order over the interval $\tau$. $\omega = 1$ represents the most favorable scenario where all cancellations result in an advancement of queue position and $\omega = 0$ the converse.*

**Remark 3.3.2** (market-by-order data). *If market-by-order data is available, then $\mathcal{D}_\tau$ contains a richer set of information that can be used to exactly determine the queue position of a cancelled limit order. Then the parameter $\omega$ can be replaced by the indicator function $\mathbb{1}_{\{\phi_{u,u} < \phi_{u,t_0}\}}$ (placed inside the summation operator) and no approximation is needed.*

It is instructive to view numerical examples of the trade-to-book ratio estimation, under either FIFO or pro-rata matching rules. These examples assume full view of the limit orders at the best buy level. Depending on the market, the size of each order may not be not known, only the total resting quantity and number of orders at the price level can typically be extracted from the market data feeds.

**Example: FIFO market** Suppose at time $t_0^-$ the queue depth at the best bid is 50. The reference limit order to buy 50 contracts at the best bid level is received by the exchange at time $t_0$. In a FIFO market, the limit order joins the back of the queue at the best bid level. Let's further suppose that between time $t_0$ and time $t$ a market sell order of size 25 subsequently arrives. Additionally, in this time interval, a best bid limit order of size 20, received prior to time $t_0$, is cancelled. As a result of the market sell order matching with 25 contracts in the best bid queue and the cancellation of the best bid order, the queue position of the reference order has advanced so that there are 5 contracts ahead of it. If a new sell market order of size 10 arrives at time $t$ then its trade-to-book ratio, with respect to the reference limit order, has the value

$$\mathcal{R}_t(50; \mathcal{D}_\tau^1, 1) = \frac{10}{50 + 50 - (25 + 1 \cdot 20 + 0)} = 2/11. \tag{9}$$

and so the reference order is partially filled. Note that the reference order is partially filled if the trade-to-book ratio is above 1/11.

**Example: Pro-Rata market** Suppose once again at time $t_0^-$ the queue depth at the best bid is 50 consisting of orders of size {25,15,10}. The reference limit order to buy 10 contracts at the best bid level is received by the exchange at time $t_0$ and joins the back of the queue in a pro-rata market. Let's further suppose that between time $t_0$ and time $t$ a market sell order of size 25 subsequently arrives and is matched with the best bid order of size 25. Additionally, in this time interval, a new best bid limit order of size 15 arrives and joins the queue in second place. The net result of the market sell order matching with 25 contracts and the new best bid order arriving is that the queue position of the reference order remains the same. However the resting quantity of contracts ahead of it has fallen from 50 to 40. If a new sell market order of size 10 arrives at time $t$ then it's trade-to-book ratio, with respect to the reference limit order, has the value

$$\mathcal{R}_t(10; \mathcal{D}_\tau^1, 1) = \frac{10}{50 + 10 - (25 + 0 - 1 \cdot 15)} = 0.2. \tag{10}$$

**Remark 3.3.3** (special cases). *In certain limiting cases, the value of $\omega$ is known. For example, in a FIFO market, if the total depth of the best bid level monotonically reduced over $\tau$ then no new orders arrived. Furthermore the change in the total depth minus the sum of market orders gives the number of orders cancelled with higher queue priority than the reference order, and hence $\omega = 1$. Again, this estimate assumes that each and every change is reported as a separate update in the market data feed.*

## 4 Market Marking Strategy

The trade-to-book-ratio will be used to simulate whether the reference order is filled. The mechanism for deciding when and how orders are placed and cancelled is given by a strategy. We restrict our consideration to market making strategies that use a prediction of the future mid-price to allocate quotes.

Suppose that at time $t_0$ a trade decision is made based on a strategy using the predicted change in mid-price over a horizon $\tau$. In the simplest case, the strategy will place a limit order at the inside market if there is no predicted mid-price change. Otherwise it will place a limit order one level away from the inside market - placing an ask above the best ask if there is a predicted up-flip or placing a bid below the best bid if there is a predicted down-flip. If a quote already exists at the inside market, the strategy will submit a cancel-replace order on prediction of a price move. The strategy can, of course, be much more elaborate using layers and other forms of trade-expression. For simplicity, we also ignore partial fills in our cash flow calculations. More formally, we define a strategy as follows:

**Definition 4.0.1** (Strategy). *A strategy is a n-vector function $\mathcal{L} : \mathbb{R}^+ \times \mathbb{Z} \cap (-m, m] \to \mathbb{Z}^n$ of the form $\mathcal{L}_t(\hat{Y}_t)$, where t denotes the time that the trade is placed. Based on the predicted value of $\hat{Y}_t$, the strategy quotes on either the bid and ask at one or more price levels.*

**Definition 4.0.2** (Market Making Strategy). *A market making strategy is the pair $\mathcal{L}_t := (\mathcal{L}_t^a, \mathcal{L}_t^b)$ representing the quoting of a bid and ask at time t.*

Note that for ease of notation, we will hereon use short-hand notation for the fill probabilities $\mathcal{R}_t^k := \mathcal{R}_t(\mathcal{L}_0^k(\hat{Y}_0); \mathcal{D}_\tau^k, \omega)$. In practice, the decision to buy or sell depends on the net position. The net position is, however, a complex function of filled current and past orders. For this reason, we introduce a state variable representing the state of whether the spread is captured, only one side is filled leading due to adverse selection or no trades on either side are filled:

**Definition 4.0.3** (Spread State). *The state of the spread at time t based on the market making strategy $\mathcal{L}_0$ is a function $Z : [-1,1] \cap \mathbb{Z} \to [-1,1] \cap \mathbb{Z}$ of the form*

$$Z_t(\hat{Y}_0) = \begin{cases} 1, & A := \bigcup_{k=1}^n \{\mathcal{R}_t^{k,a} \geq 1\} \cap \bigcup_{k=1}^n \{\mathcal{R}_t^{k,b} \geq 1\} \neq \emptyset, \\ -1, & B := \bigcup_{k=1}^n \{\mathcal{R}_t^{k,a} < 1\} \cap \bigcup_{k=1}^n \{\mathcal{R}_t^{k,b} < 1\} \neq \emptyset, \\ 0, & (A \cup B)^c \neq \emptyset. \end{cases} \quad (11)$$

**Definition 4.0.4** (Realized P&L). *Let $\Phi : [-1,1] \cap \mathbb{Z} \to \mathbb{R}$ denote the realized P&L from capturing the spread or adverse selection, after including transactions costs c :*

$$\Phi(z) = \begin{cases} \mathcal{L}^a(\hat{Y}_t) \cdot \mathbf{s}_t^a - \mathcal{L}^b(\hat{Y}_t) \cdot \mathbf{s}_t^b - 2Lc, & z = 1, \\ \mathcal{L}^a(\hat{Y}_t) \cdot \mathbf{s}_t^a - \mathcal{L}^b(\hat{Y}_t) \cdot \mathbf{s}_t^b - L(\delta + 2c), & z = 0. \end{cases} \quad (12)$$

*The size of the order on each side of the book is assumed to be the same $|\mathcal{L}^a| = |\mathcal{L}^b| = L$ and c is the transaction cost per contract.*

**Remark 4.0.5** (Adverse selection cost). *Implicit in our assumption of cash flow is that an adverse selection leads to a relative loss of one tick per contract as one side is filled at a competitive price and the market maker must then quote a less competitive price on the other side to close out the position. In some cases, as for example in strategy $s_1$, this leads to transaction costs without capturing the spread.*

The cash flow at time $t$ from the market making strategy $\mathcal{L}_0 := (\mathcal{L}_0^a, \mathcal{L}_0^b)$ as a function of the prediction $\hat{Y}_0$ is given by

$$V_t(\hat{Y}_0) = \sum_{z \in \{0,1\}} \mathbf{1}_{\{Z_t(\hat{Y}_0) = z\}} \Phi(z). \quad (13)$$

whose value depends on the data $\mathcal{D}_\tau$ and the triple $(\mathcal{L}, \hat{Y}_0, Y_0)$, consisting of all offers placed by the strategy $\mathcal{L}_0$, the prediction $\hat{Y}_0$ and the true state $Y_0$ at time $t_0$.

**Example: Market Marking Strategy** An example of a simple market making strategy $\mathcal{L}_0 = (\mathcal{L}_0^a, \mathcal{L}_0^b)$, assuming $n = 2$, which either quotes at or away from either side of the inside market is

$$\mathcal{L}^a(\hat{Y}_0) \begin{cases} \{0, L\}, & \hat{Y}_0 = 1, \\ \{L, 0\}, & \hat{Y}_0 = 0, \\ \{L, 0\}, & \hat{Y}_0 = -1. \end{cases} \quad (14) \qquad \mathcal{L}^b(\hat{Y}_0) \begin{cases} \{L, 0\}, & \hat{Y}_0 = 1, \\ \{L, 0\}, & \hat{Y}_0 = 0, \\ \{0, L\}, & \hat{Y}_0 = -1. \end{cases} \quad (15)$$

$$\Phi(z) = \begin{cases} L\left(\delta n(\hat{Y}_0) - c\right), & 1 \\ L\left(\delta(n(\hat{Y}_0) - 1) - c\right), & 0 \end{cases} \quad (16)$$

where $c$ is a round-trip transaction cost and $n : [-1,1] \cap \mathbb{Z} \to [1,2] \cap \mathbb{Z}$ with $n(0) = 1$ and $n(-1) = n(1) = 2$. The cash flow at time $t$ from the strategy $\mathcal{L}_0$ as a function of the prediction $\hat{Y}_0$ is given by

$$V_t(\hat{Y}_0) = \sum_{z \in \{0,1\}} \mathbf{1}_{\{Z_t(\hat{Y}_0) = z\}} \Phi(z). \quad (17)$$

**Remark 4.0.6** (latency). *Our notation may suggest that order receipt time and decision time are the same. In practice, latency l between the participant and the exchange is an important factor and a trade decision must be made at time $t_0 - l$ for it to be received at time $t_0$. Additional constraints are usually added to the strategy to manage inventory but have been ignored here for ease of exposition.*

## 4.1 Strategy Performance

The assessment of the economic impact of a trade decision can be measured as a function of error in the prediction. We recall the standard definition of a confusion matrix used to characterize error.

**Definition 4.1.1** (Confusion Matrix). *The confusion matrix is a function $C : \mathbb{R}^+ \to \mathbb{R}_+^{M \times M}$, $M = 2m+1$ of the form*

$$C_{ij}(t) := P(\hat{Y}_t = y_j \mid Y_t = y_i), \forall i, j \in \{1, \ldots, M\} \times \{1, \ldots, M\}, \tag{18}$$

*for a predicted state $\hat{Y}_t \in \mathbf{y} := \in [-m, m] \cap \mathbb{Z}$ and a true state $Y_t \in \mathbf{y}$.*

The economic impact of a level $k$ quote placed at time $t_0$ can be measured by its expected cash flows conditioned on the pair $(Y_0, \hat{Y}_0)$. We use a matrix to represent the expected cash flows from this quote over all combinations of true-predicted value pairs.

**Definition 4.1.2** (Trade Information Matrix). *The trade information matrix is a function $T : \mathbb{R}^+ \to \mathbb{R}^{M \times M}$, $M = 2m+1$ of the form*

$$T_{ij}(t; \Omega_0^k, \mathcal{D}_\tau^k, \omega) := P(Y_0 = y_i)\mathbb{E}[V_t(\hat{Y}_0 = y_j)|Y_0 = y_i, \hat{Y}_0 = y_j] \tag{19}$$

*which uses the triple $\Omega_0^k := (\mathcal{L}_0^k, \hat{Y}_0, Y_0)$, consisting of predictions $\hat{Y}_0$, the true state $Y_t$ and the $k^{th}$ level offer placed by a strategy $\mathcal{L}_0$ at time $t_0$, in addition to the order book events $\mathcal{D}_\tau^k$.*

**Theorem 4.1.3** (Expected Cash Flow). *The expected cash flow from the triple $\Omega_0 := (\mathcal{L}_0, \hat{Y}_0, Y_0)$ is*

$$\mathbb{E}[V_t] = tr(C(t_0)T'(t)) \tag{20}$$

*where $T'$ denotes the transpose of $T$.*

*Proof.* From Equation 13, the expected value of the cash flow is

$$\mathbb{E}[V_t] = \sum_{z \in \{0,1\}} P(Z_t(\hat{Y}_0) = z)\Phi(Z_t = z) \tag{21}$$

Using the following property of conditional probabilities

$$P(Z_t = z) = \sum_i P(Z_t = z|Y_0 = y_i)P(Y_0 = y_i) \tag{22}$$

$$= \sum_j \sum_i P(Z_t = z|Y_0 = y_i, \hat{Y}_0 = y_j)P(\hat{Y}_0 = y_j|Y_0 = y_i)P(Y_0 = y_i) \tag{23}$$

and inserting the far right hand side of the above expression into Equation 21 gives

$$\mathbb{E}[V_t] = \sum_j \sum_i \sum_{z \in \{0,1\}} P(Z_t = z|Y_0 = y_i, \hat{Y}_0 = y_j)P(\hat{Y}_0 = y_j|Y_0 = y_i)P(Y_0 = y_i)\Phi(z) \tag{24}$$

$$= \sum_j \sum_i \mathbf{E}[V_t(\hat{Y}_0 = y_j)|Y_0 = y_i, \hat{Y}_0 = y_j]P(\hat{Y}_0 = y_j|Y_0 = y_i)P(Y_0 = y_i) \tag{25}$$

Inserting the definition of the confusion matrix and rearranging the order of the terms

$$\mathbb{E}[V_t] = \sum_i \sum_j \sum_{z \in \{0,1\}} C_{ij}(t_0)P(Y_0 = y_i)\mathbf{E}[V_t(\hat{Y}_0 = y_j)|Y_0 = y_i, \hat{Y}_0 = y_j] \tag{26}$$

and from the definition of $T_{ij}(t)$ gives

$$\mathbb{E}[V_t] = tr\left(C(t_0)T'(t)\right). \tag{27}$$

□

**Remark 4.1.4** (Fill probability estimation). *The magnitude of the entries in the trade information matrix depend on the fill probabilities, conditioned on true state, the probability of a change in state and the prediction signal based strategy. The fill probabilities are the most difficult to estimate because, as previously mentioned, they depend on a number of factors including the matching engine algorithm, order book imbalance and queue priority. The following example assumes a parameter form for these probabilities, however, as described later in the paper, it is assumed that they are estimated from an exchange simulator.*

**Example: Parametric fill probabilities** We now illustrate the trade information matrix by comparing two strategies. One strategy $s_1 = (\{1,0\},\{1,0\})$ simply places a one lot bid at the inside market and does not use a prediction. The other strategy $s_2 = (\mathcal{L}_0^a, \mathcal{L}_0^b)$ does use the prediction $\hat{Y}_0$ as follows:

$$\mathcal{L}^a(\hat{Y}_0) = \begin{cases} \{0,1\}, & \hat{Y}_0 = 1, \\ \{1,0\}, & \hat{Y}_0 = 0, \\ \{1,0\}, & \hat{Y}_0 = -1. \end{cases} \qquad \mathcal{L}^b(\hat{Y}_0) = \begin{cases} \{1,0\}, & \hat{Y}_0 = 1, \\ \{1,0\}, & \hat{Y}_0 = 0, \\ \{0,1\}, & \hat{Y}_0 = -1. \end{cases}$$

To illustrate the evaluation of the expected cash flows with the trade information matrix, assume now that conditional fill probabilities and the conditional probability of a change of mid-price are given by the exponential distributions in Table 3. $\lambda_1$ is the decay parameter for the mid-price change probabilities. $\lambda_2$ is the decay parameter for the fill probabilities. The parameter $a > 0$ is used to control the asymmetry in the mid-price movements, where $a = 0.5$ is the symmetric case. The parameter $b$ is used to adjust the relative fill rates by level. It is assumed that quotes at outer prices levels are more difficult to be filled and hence $0 < b < 1$.

|  | $y$ | | |
|---|---|---|---|
|  | -1 | 0 | 1 |
| $P(Y_0 = y)$ | $a\left(1 - \exp(-\lambda_1(t - t_0))\right)$ | $\exp(-\lambda_1(t - t_0))$ | $(1 - a)\left(1 - \exp(-\lambda_1(t - t_0))\right)$ |
| $P(\mathcal{R}_t^{a,1} \geq 1 \mid Y_0 = y)$ | $1 - \exp(-\lambda_2 b(t - t_0))$ | $1 - \exp(-\lambda_2(t - t_0))$ | 1 |
| $P(\mathcal{R}_t^{b,1} \geq 1 \mid Y_0 = y)$ | 1 | $1 - \exp(-\lambda_2(t - t_0))$ | $1 - \exp(-\lambda_2 b(t - t_0))$ |
| $P(\mathcal{R}_t^{a,2} \geq 1 \mid Y_0 = y)$ | 0 | 0 | $1 - \exp(-\lambda_2(t - t_0))$ |
| $P(\mathcal{R}_t^{b,2} \geq 1 \mid Y_0 = y)$ | $1 - \exp(-\lambda_2(t - t_0))$ | 0 | 0 |

Table 3: *This table shows the conditional fill probabilities and the mid-price change probabilities. $\lambda_1$ is the decay parameter for the mid-price change probabilities. $\lambda_2$ is the decay parameter for the fill probabilities. The parameter $a > 0$ is used to control the asymmetry in the mid-price movements, where $a = 0.5$ is the symmetric case. The parameter $b$ is used to adjust the relative fill rates by level. It is assumed that quotes at outer prices levels are more difficult to be filled and hence $0 < b < 1$.*

Figure 3 shows the expected realized P&L of strategy $s_2$ (red line) compared with strategy $s_1$ (black line) under the probability model defined in Table 3 with the following parameterization $a = 0.5, b = 0.5, \lambda_1 = 1$. The spread $\delta = \$12.5$ and the round-trip transaction cost is $c = \$0.7$. Larger values of $\lambda_2$ correspond to increased growth rates of the fill probabilities. The figure also compares the effect of a white noise predictor (dotted) with a perfect predictor (dashed). The equilibrium level of the white noise predictor is lower than the perfect predictor but still more profitable than strategy $s_1$.

Table 4 show the trade information matrices for the $s_1$ (left) and $s_2$ (right) strategies. The elements in the trade information matrix for $s_1$ are the same in each row because the strategy is stateless. Therefore the expected P&L is independent of the confusion matrix. We observe that most of the P&L is generated when there is no price-flip although small, and equal, amounts are generated when there is either a price up flip or down flip. Note that the top and bottom rows are smaller in magnitude reflecting the relatively low probability of a price-flip compared to a stationary mid-price.

The elements in each row of the trade information matrix for $s_2$ are not always the same. If the price is stationary but the strategy receives a non-zero signal, then it will quote away from the inside market with any higher gain due to the increased potential for capturing a 2 tick spread exactly offset by the lower probability of a fill. And so, in this special case, the outcome of the prediction was irrelevant to the gain

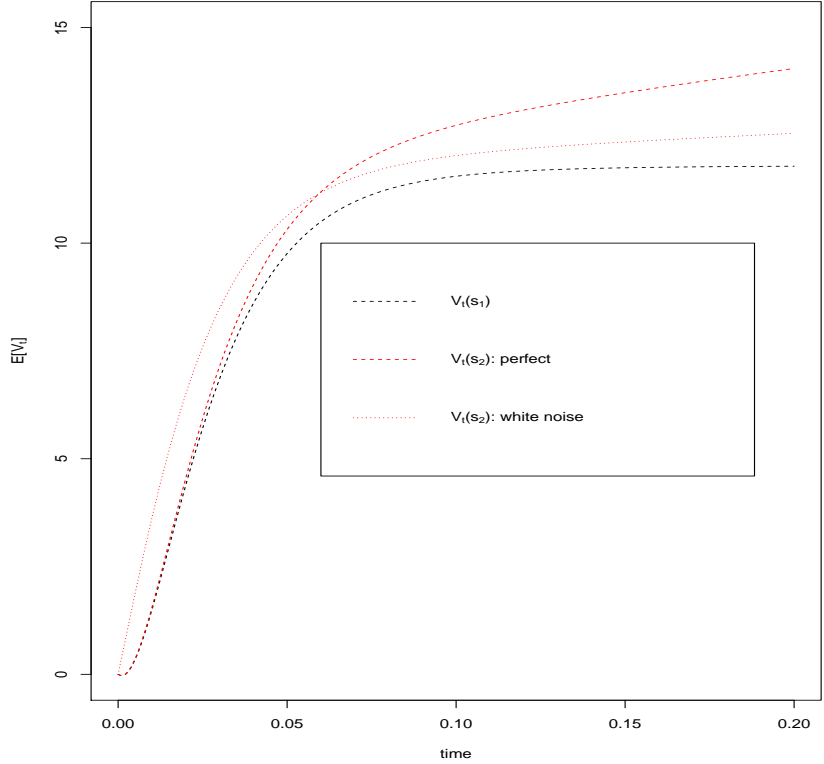

Figure 3: *This figure shows the expected realized P&L of strategy $s_2$ compared with strategy $s_1$ for the following configuration $a = 0.5, b = 0.5, \lambda_1 = 1$. The spread $\delta = \$12.5$ and the round-trip transaction cost is $c = \$0.7$. Larger values of $\lambda_2$ correspond to increased growth rates of the fill probabilities. The figure also compares the effect of a white noise predictor (dotted) with a perfect predictor (dashed). The equilibrium level of the white noise predictor is lower than the perfect predictor but more profitable than strategy $s_1$.*

- the strategy is robust to false positives or negatives when the true state is stationary. When there is a price-flip, the most profit is gained with a correct prediction.

|    | $\hat{Y}_0 = y$ | | |    | $\hat{Y}_0 = y$ | | |
|----|-------|-------|-------|----|--------|-------|--------|
|    | -1    | 0     | 1     |    | -1     | 0     | 1      |
| -1 | 1.062 | 1.062 | 1.062 | -1 | 2.195, | 1.062 | 1.0695 |
| 0  | 9.660 | 9.660 | 9.660 | 0  | 9.660  | 9.660 | 9.660  |
| 1  | 1.062 | 1.062 | 1.062 | 1  | 1.0695 | 1.062 | 2.195  |

Table 4: *These tables show the trade information matrices for the $s_1$ (left) and $s_2$ (right) strategies evaluated at elapsed time $t = 0.2$.*

The above example provides some insight into the conditions under which a market making strategy that exploit a prediction signal will outperform a naive strategy. The remainder of the paper shall now describe the estimation of the trade information matrix from market data, using empirical probabilities rather than parametric, and further evaluate the comparative gains of a market making strategy. In particular we estimate the fill and mid-price change probabilities using our exchange simulator.

## 4.2 High Frequency Data

Our dataset is an archived Chicago Mercantile Exchange (CME) FIX format message feed captured from August 1, 2016 to August 31, 2016. This message feed records all transactions in the front month E-mini S&P 500 futures contract (ESU6) between the times of 12:00pm and 22:00 UTC. The ES tick size is a quarter of a point, or 12.50 per contract (rounded to the nearest cent). We extract details of each limit order book update, including the nano-second resolution time-stamp, the quoted price and depth for each limit order book level.

The mid-price at time $t$ is denoted by

$$p_t = \frac{s_t^{a,1} + s_t^{b,1}}{2}. \tag{28}$$

This mid-price can evolve in minimum increments of half a tick but is almost always observed to move at increments of a tick over time intervals of a milli-second or less. In our feature set, each limit order book update is recorded as an observation. Each observation is labelled based on whether the mid-price will increase, decrease or remain constant over a horizon $h$:

$$Y_t = \Delta p_{t+h}^t, \tag{29}$$

where $\Delta p_{t+h}^t$ is the forecast of discrete mid-price changes from time $t$ to $t+h$, given measurement of the predictors up to time $t$. The forecasting horizon $h$ can be chosen to represent a fixed number of events or can be a fixed time interval. This choice is based on practical considerations which are discussed later in Section 5.

Table 5 shows the limit order book before and after the arrival of the sell aggressor. Here, the response $Y_t$ is mid-price movement, in units of ticks, between the current and next tick.

| Timestamp | $s_t^{b,1}$ | $s_t^{b,2}$ | ... | $q_t^{b,1}$ | $q_t^{b,2}$ | ... | $s_t^{a,1}$ | $s_t^{a,2}$ | ... | $q_t^{a,1}$ | $q_t^{a,2}$ | ... | $Y_t$ |
|---|---|---|---|---|---|---|---|---|---|---|---|---|---|
| 06:00:00.015 | 2175.75 | 2175.5 | ... | 103 | 177 | ... | 2176 | 2176.25 | ... | 82 | 162 | ... | -1 |
| 06:00:00.036 | 2175.5 | 2175.25 | ... | 177 | 132 | ... | 2175.75 | 2176 | ... | 23 | 82 | ... | 0 |

Table 5: *This table shows the limit order book of ESU6 before and after the arrival of the sell aggressor listed in Figure 2. Here, the response is the mid-price movement over the subsequent interval, in units of ticks. $s_t^{b,i}$ and $q_t^{b,i}$ denote the level $i$ quoted bid price and depth of the limit order book at time $t$. $s_t^{a,i}$ and $q_t^{a,i}$ denote the corresponding level $i$ quoted ask price and depth.*

The result of categorizing (a.k.a. labeling) each observation leads to a class imbalance problem as, at such a short prediction horizon, approximately 99% of the observations have a zero response. To construct a 'balanced' training set, observations (sequences of input variables) labeled by the minority class are oversampled with replacement and the majority class observations are undersampled without replacement.

Following Kercheval and Zhang (2015); Sirignano (2016); Dixon (2017), we compose our feature set of five levels of prices, volumes and number of limit orders on both the ask and bid side of the book. We additionally, and somewhat heuristically via a process of 'feature engineering', characterize order flow by the ratio of the number of market buy orders arriving in the prior 50 observations to the resting number of ask limit orders at the top of book. We estimate the analogous ratio for the sell by orders. This rationale for this ratio is motivated by our observation that an increase in this ratio will more likely deplete the best ask level and the mid-price will up-tick, and vice-versa for a down-tick. The combination of this spatial representation of the limit order book and the order flow gives a total of $P = 32$ features.

## 5 Results

The exact architecture and weight matrix sizes of our recurrent neural network are given by

$$\text{output}: Y^k = \text{softmax}(F_y^k(h_T)) = \frac{\exp(F_y^k(h_T))}{\sum_{j=1}^K \exp(F_y^j(h_T))},$$

$$\text{hidden states}: h_t = \max(W_h X_t + U_h h_{t-1} + b_h, 0), \quad t = 1, \ldots, T,$$

where $W_h \in \mathbb{R}^{20 \times 32}, U_h \in \mathbb{R}^{20 \times 20}$ and $W_y \in \mathbb{R}^{3 \times 20}$. We initialize the hidden states to zero. We use the SGD method, implemented in `Python`'s `TensorFlow` Abadi et al. (2016) framework, to find the optimal network weights, bias terms and regularization parameters. We employ an exponentially decaying learning rate schedule with an initial value of $10^{-2}$. The optimal $\ell_2$ regularization is found, via a grid-search, to be $\lambda_2 = 0.01$. The Glorot and Bengio method is used to initialize the weights of the network Glorot and Bengio (2010).

Times series cross-validation is performed over 20 consecutive trading days. Training sets are compiled from the previous 3 trading days and contain, on average, $5,192,822$ observations. These sets are balanced resulting in a reduced training set size of typically just less than $100,000$ observations. The validation and test sets are compiled for the next trading day following the 3 day training period. These are unbalanced, with the verification set containing $2 \times 10^5$ observations and the remaining test set, on average, containing approximately $1.5 \times 10^6$ observations. Each experiment is run for 1000 epochs with a mini-batch size of 500 drawn from the training set of 32 input variables. We follow the standard convention of choosing the number of epochs based on convergence of the cross-entropy and the mini-batch size is chosen for computational performance. Each sequence is chosen to be of length 10 and the number of hidden units is chosen between 10 and 20. The gridded search to find the optimal network architecture and regularization parameters takes several hours on a modern graphics processing unit (GPU). The search yields several candidate architectures and parameter values.

To reliably measure performance on the unbalanced test set, we compute the $F1$ score - the geometric mean of the precision and recall:

$$F1 = 2 \frac{\text{precision} \cdot \text{recall}}{\text{precision} + \text{recall}}.$$

The $F1$ score is designed for binary classification problems. When the data has more than two classes, the $F1$ score is provided for each class $k$ by setting $Y = k$ to the positive and all remain classes to the negative label. Table 6 compares the performance of the RNN with the elastic net method, implemented in the R package `glmnet` (Friedman et al., 2010; Simon et al., 2011), for predicting the next price movement. The elastic-net method, with $\alpha = 0.5$, exhibits an out-of-sample classification accuracy of 49.6%. Table 7 compares the confusion matrices for the (left) in-sample performance of the RNN classifier with the (right) out-of-sample performance of the RNN classifier using a prediction horizon $h = 1ms$. Each confusion matrix is formed by averaging the daily confusion matrices over the month of August. The parentheses show the standard deviation.

| Model | F1 (-1) | F1(0) | F1 (1) | Accuracy |
|---|---|---|---|---|
| Elastic-Net | 0.116 (0.061) | 0.649 (0.156) | 0.108 (0.055) | 0.626 (0.141) |
| RNN | 0.165 (0.050) | 0.881 (0.050) | 0.161 (0.045) | 0.817 (0.187) |

Table 6: *This table summarizes the performance of the RNN over a 20 trading day test period. The RNN classifier is used to predict the categories: $\{Y = -1, Y = 0, Y = 1\}$. The mean and standard deviation of the f1 scores are compared with the elastic-net method.*

| | In Sample | | | | Out-of-Sample | | |
|---|---|---|---|---|---|---|---|
| | -1 | 0 | 1 | | -1 | 0 | 1 |
| -1 | 0.93 (0.03) | 0.03 (0.01) | 0.04 (0.03) | -1 | 0.84 (0.06) | 0.06 (0.05) | 0.10 (0.04) |
| 0 | 0.10 (0.02) | 0.81 (0.03) | 0.09 (0.02) | 0 | 0.10 (0.03) | 0.78 (0.06) | 0.12 (0.03) |
| 1 | 0.05 (0.02) | 0.04 (0.02) | 0.91 (0.02) | 1 | 0.11 (0.04) | 0.06 (0.04) | 0.83 (0.04) |

Table 7: *This table compares the confusion matrices for the (left) in-sample performance of the RNN classifier with the (right) out-of-sample performance of the RNN classifier using a prediction horizon $h = 1ms$. Each confusion matrix is formed by averaging the daily confusion matrices over the month of August. The parentheses show the standard deviation.*

## 5.1 Strategy comparison

The estimation of the trade information matrix from a market simulator is now described. A new prediction is made for each limit order book in the market data. By using the market simulator to evaluate the current position, a market making strategy will typically quote to maintain a target position and remain within an inventory constraint. A prediction based market marking strategy will additionally adjust the quote if the prediction changes, even if the overall position hasn't changed since the previous book update. In our experiments, we update the quotes at the end of each prediction horizon which has been chosen to be $h = 1s$. Our trade execution model assumes that $\omega = 0$, leading to the most conservative queue position estimate.

We first illustrate the two market making strategies, given in Example 4.1, with a backtest simulation. Figure 4 (top) compares the P&L of strategy $MM1$, the prediction agnostic strategy, to the prediction based strategy $MM2$. Each strategy is applied to ESU6 data from 11am to 12:30pm (CST) on August the 4th, 2016 and a unit inventory constraint is set. The classifier is trained from the market data using the previous three trading days and a 1s prediction horizon. With perfect information, and 0ms latency, strategy $MM2$ is shown to outperform strategy $MM1$. The backtest assumes that the market maker is a clearing member or equity member firm with a transaction cost per contract of 35 cents CME (2017).

Each strategy modifies its execution to ensure that the overall position remains within a limit. Figure 4 (bottom) shows the position over time resulting from the MM2 strategy with a unit position limit.

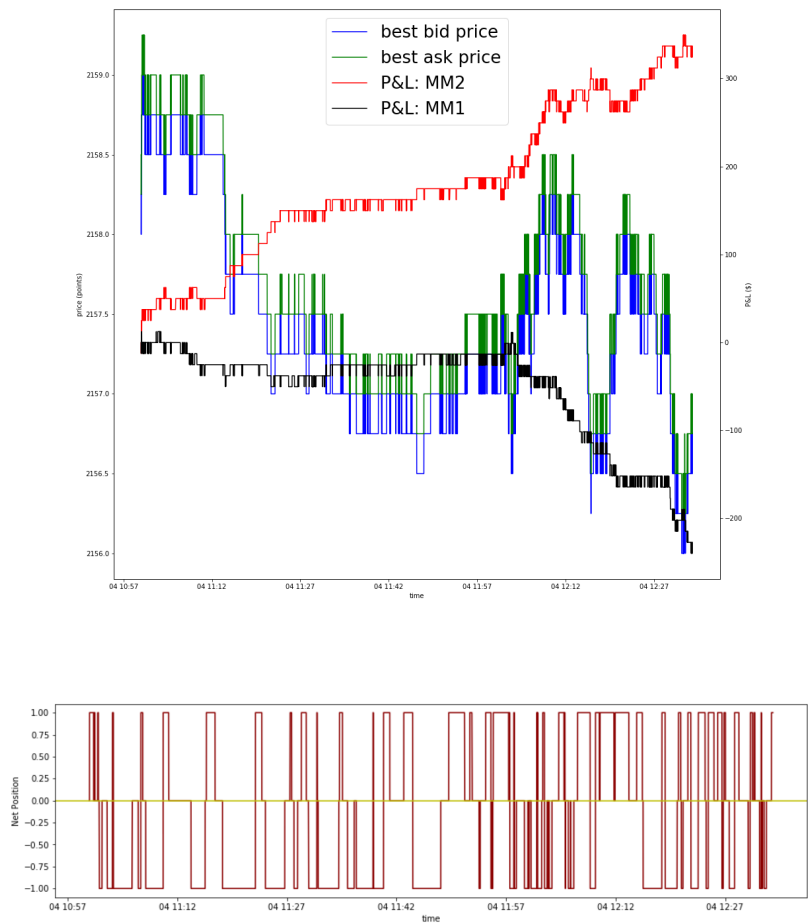

Figure 4: *(top) The top-of-the book prices together with the cumulative P&L of the market making strategies are shown over the period 11am to 12:30pm (CST) on August the 4th, 2016. (bottom) The position in ESU6 resulting from the MM2 strategy with a unit position limit.*

Figure 5 (left) shows the empirical probabilities of price-flips against prediction horizon $h$. The probabilities of price-flips are similar and observed to increase to approximately 0.17 at a horizon of 1s shows the empirical distribution of fill times (ms) on August 4th 2016. The boxplot in the right hand panel shows the empirical distribution of fill times. After 1 second, the order is cancelled if it is no longer at the target price level. Fill times are separately estimated for best bid and ask quotes conditioned on $Y_0$. When the book is stationary over the next interval, the distribution of best bid fill times is similar to the best ask fill times. The median of the best ask fill times is lower when there is a price-up-flip since by construction the best ask level will, modulo the reference order, be cleared. Conversely, when there is a price-down-flip, the best bid level will be cleared. The best bid and best ask can be filled even if the mid-price movement is away from those levels, but it is much rarer and tends to occurs almost immediately and well before the price movement.

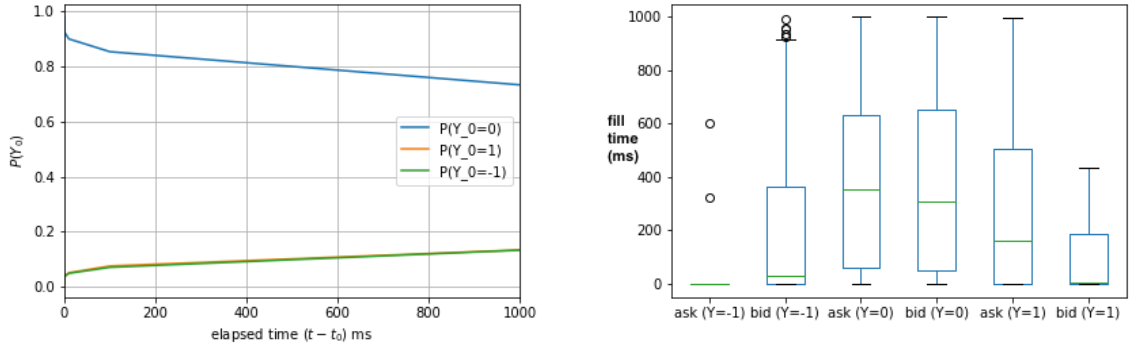

Figure 5: *(left) The empirical probabilities of price-flips are shown against prediction horizon $h$. The probabilities of price-flips are similar and observed to increase to approximately 0.17 at a horizon of 1s. The boxplot (right) shows the empirical distribution of fill times on August 4th 2016. Empirical fill times are separately estimated for best bid and ask quotes conditioned on $Y_0$. When the book is stationary over the next interval, the distribution of best bid fill times is similar to the best ask fill times. The median of the best ask fill times is lower when there is a price-up-flip since, by construction, the best ask level is, modulo the reference order, cleared. Conversely, when there is a price-down-flip, the best bid level is cleared. The best bid and best ask are filled even if the mid-price moves away from those levels, but it is much rarer and tends to occurs almost immediately and well before the price movement.*

Table 8 shows the estimated empirical price movement probabilities, quote fill probabilities and spread fill probabilities conditioned on the movement of the true state over a forecasting horizon of $t = h = 1s$. Each column shows the corresponding conditional probabilities for each value of $Y_0$. These probabilities are estimated using the market simulator applied to ESU6 data between 6am and 4pm CST on August 4th, 2016.

|  | $y$ | | |
| --- | --- | --- | --- |
|  | -1 | 0 | 1 |
| $P(Y_0 = y)$ | 0.134 | 0.732 | 0.134 |
| $P(R_t^{a,0} \geq 1 \mid Y_0 = y)$ | 0.074 | 0.67 | 0.581 |
| $P(R_t^{b,0} \geq 1 \mid Y_0 = y)$ | 0.563 | 0.615 | 0.107 |
| $P(Z = 1 \mid Y_0 = y)$ | 0.042 | 0.412 | 0.062 |
| $P(Z = 0 \mid Y_0 = y)$ | 0.554 | 0.461 | 0.563 |

Table 8: *This table shows the estimated empirical price movement probabilities, quote fill probabilities and spread fill probabilities conditioned on the movement of the true state over a forecasting horizon of $t = h = 1s$. Each column shows the corresponding conditional probabilities for each value of $Y_0$.*

Table 9 show the trade information matrices for the MM1 (left) and MM2 (right) strategies. The elements in the trade information matrix for MM1 are the same in each row because the strategy is stateless. Therefore the expected P&L is independent of the confusion matrix. We observe that most of the P&L is generated when there is no price-flip although small, but not equal, amounts are generated when there is either a price up flip or down flip. Note that the top and bottom rows are smaller in magnitude reflecting the relatively low probability of a price-flip compared to a stationary mid-price.

The elements in each row of the trade information matrix for MM2 are not the same. If the price is stationary but the strategy receives a non-zero signal, then it will quote away from the inside market with a higher gain due to the increased potential for capturing a 2 tick spread.

The elements in each row of the trade information matrix for $MM2$ are not the same. If the price is stationary but the strategy receives a non-zero signal, then it will quote away from the inside market with any higher gain due to the increased potential for capturing a 2 tick spread being unevenly offset by the lower probability of a fill. For the stationary case, we observe that more profit can be gained from a wrong prediction. Even though the spread is not filled if the prediction is wrong, adverse selection may still result in a higher gain by quoting away from the inside market than quoting at the inside market. And so the strategy is robust to false positives or negatives when the true state is stationary (and may actually benefit from them if the adverse selection gains are sufficiently favorable). When there is a price-flip, profit is only gained with a correct prediction.

|    | $\hat{Y}_0 = y$ |       |       |    | $\hat{Y}_0 = y$ |       |       |
|----|------|-------|-------|----|-------|-------|-------|
|    | -1   | 0     | 1     |    | -1    | 0     | 1     |
| -1 | 0.014| 0.014 | 0.014 | -1 | 0.604 | 0     | 0     |
| 0  | 3.324| 3.324 | 3.324 | 0  | 2.457 | 2.225 | 4.059 |
| 1  | 0.045| 0.045 | 0.045 | 1  | 0     | 0     | 0.640 |

Table 9: *These tables show the trade information matrices for the MM1 (left) and MM2 (right) strategies. table shows the trade information matrix for all quotes placed at the inside market and the bottom table corresponds to quotes placed at the next price level away from the inside market.*

Further insight into the values of the trade information matrices can be gained by studying the constituent probabilities. Table 10 shows the estimated empirical price movement probabilities, Level 1 and 2 quote and spread fill probabilities conditioned on the movement of the true state over a forecasting horizon of 1$s$. Each column shows the corresponding conditional probabilities for each value of $Y_0$. These probabilities are estimated using the market simulator applied to ESU6 data between 6am and 4pm CST on August 4th, 2016. Table 11 shows the probability that the spread is filled (top) and the probability of adverse selection using the MM2 strategy conditioned on the true movement $Y_0$ and the prediction $\hat{Y}_0$.

|    |    | $y$ |    |
|----|----|----|----|
|    | -1 | 0 | 1 |
| $P(Y_0 = y)$ | 0.134 | 0.732 | 0.134 |
| $P(R_t^{a,1} \geq 1 | Y_0 = y)$ | 0.007 | 0.511 | 0 |
| $P(R_t^{a,2} \geq 1 | Y_0 = y)$ | 0 | 0 | 0.421 |
| $P(R_t^{b,1} \geq 1 | Y_0 = y)$ | 0 | 0.504 | 0.011 |
| $P(R_t^{b,2} \geq 1 | Y_0 = y)$ | 0.403 | 0 | 0 |

Table 10: *This table shows the estimated empirical price movement probabilities, Level 1 and 2 quote and spread fill probabilities conditioned on the movement of the true state over a forecasting horizon of 1$s$. Each column shows the corresponding conditional probabilities for each value of $Y_0$.*

| $P(Z_t = 1\|Y_0 = y_i, \hat{Y}_0 = \hat{y}_j)$ | | | | $P(Z_t = 0\|Y_0 = y_i, \hat{Y}_0 = \hat{y}_j)$ | | | |
|---|---|---|---|---|---|---|---|
| | $\hat{y}$ | | | | $\hat{y}$ | | |
| $y$ | -1 | 0 | 1 | $y$ | -1 | 0 | 1 |
| -1 | 0.003 | 0 | 0 | -1 | 0.404, | 0.003 | 0 |
| 0 | 0 | 0.258 | 0 | 0 | 0.305 | 0.5 | 0.504 |
| 1 | 0 | 0 | 0.005 | 1 | 0 | 0.011 | 0.423 |

Table 11: *This table shows the probability that the spread is filled (top) and the probability of adverse selection using the MM2 strategy conditioned on the true movement $Y_0$ and the prediction $\hat{Y}_0$.*

**Remark 5.1.1** (Backtesting challenges). *The fill probability estimates highlight a complex design decision in the market simulator because the mere placement of an order changes the future state of the book in a way that could not be known with certainty at decision time. Therefore a future price-flip does not guarantee that a quote at best ask will be filled. For example, suppose at $t_0$ the depth of the best ask level includes two orders, one of which has been placed by another participate and one is our reference order. The labeling of a price-flip does not account for our reference order. And so in the event that the market participate cancels the order or it is filled by a matching market order, the data is labelled as an up-flip. But the presence of our order, had the label accounted for such a position, would not necessarily have been filled. This problem arises because we have chosen to separate the labeling from the strategy.*

*It is however possible to quantify this effect by estimating the expected value of a trade if the order was always inconsequential to the sequence of price-flips. Here, we estimate the expected value under the assumption that the fill probabilities of a best bid or best ask offer are unity under a down-flip or up-flip respectively. The true state of the order book in response to a reference order can not be known of course. This is the reason why many market making firms use live simulation to accurately measure the strategy performance rather than historical simulation. Our goal here, however, is to measure the comparative advantages of a strategy against a baseline in the presence of decision error. Our approach is not intended to replace live simulation, or for that matter historical simulation, only serve as a computationally tractable approach for understanding the implications of decision error.*

For the purpose of assessing how error affects the expected realized profit, we shall assume that the confusion matrix is symmetric and parameterized by $\epsilon \in [0, 2/3]$:

$$C_{ij} = \begin{cases} 1 - \epsilon, & i = j, \\ \frac{\epsilon}{2}, & i \neq j. \end{cases} \qquad (30)$$

$\epsilon = 0$ represents a perfect predictor and $\epsilon = 2/3$ represents uniform white noise. Figure 6 compares the expected P&L of the two market market strategies as a function of error $\epsilon$ in the confusion matrix. It is observed that the expected realized profit from MM2 (red) linearly decays with $\epsilon$, to the extent that it can become less profitable than the baseline strategy MM1 (black). The figure suggests that our confusion matrices estimated in the previous section are sufficiently accurate to warrant market making with a prediction.

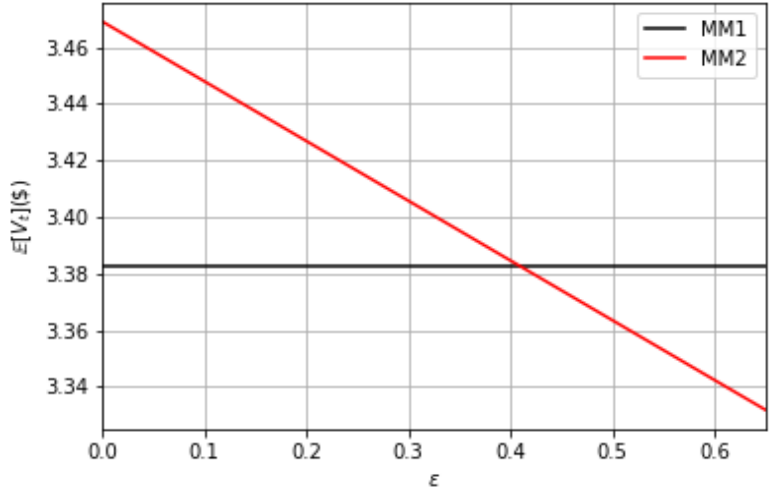

Figure 6: *This figure compares the expected P&L of the two market market strategies as a function of error $\epsilon$ in the confusion matrix. It is observed that the expected realized profit from MM2 (red) linearly decays with $\epsilon$, to the extent that it can become less profitable than the baseline strategy MM1 (black).*

## 5.2 Classifier error characteristics

In this section, we provide further assessment of the RNN performance over the month of August 2016. Figure 7 shows the F1 scores of the RNN, measured on any hourly basis. We observe little variation in the F1 scores for up-tick and down-tick predictions and hence no evidence of secular F1 score decay and hence the need to retrain the model intra-day. In fact when we observe the opposite from the F1 score for predicting no book movement- the F1 score is observed to gradually increase over the course of the day. Figure 8 shows the importance of retraining the model on a daily basis. The F1 scores are observed to exhibit a secular decay when the model is not re-trained on a daily basis but instead trained only at the beginning of the month. While the evidence for secular decay is clear, it is surprising to observe that a model trained at the beginning of the month can still be effective weeks later. Typically econometrics models for non-stationary time series perform very poorly when not frequently re-fitted to new data. This surprising model robustness, without re-training, suggests a robust relationship between the liquidity imbalance and price movements.

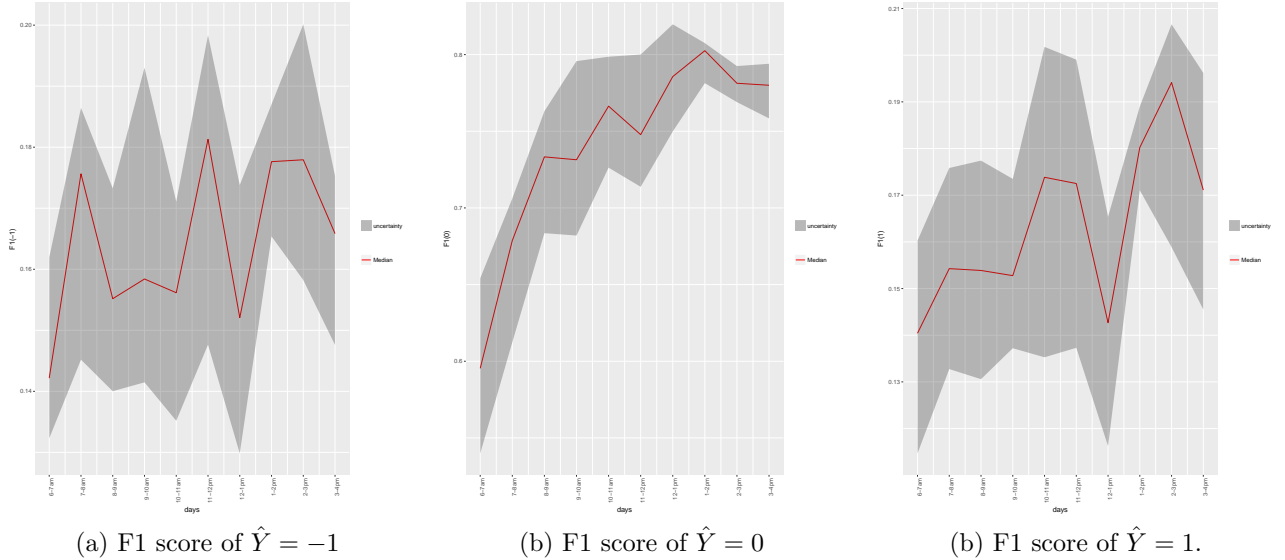

(a) F1 score of $\hat{Y} = -1$    (b) F1 score of $\hat{Y} = 0$    (b) F1 score of $\hat{Y} = 1$.

Figure 7: *The intra-day F1 scores are shown for (left) downward, (middle) neutral, or (right) upward next price movement prediction.*

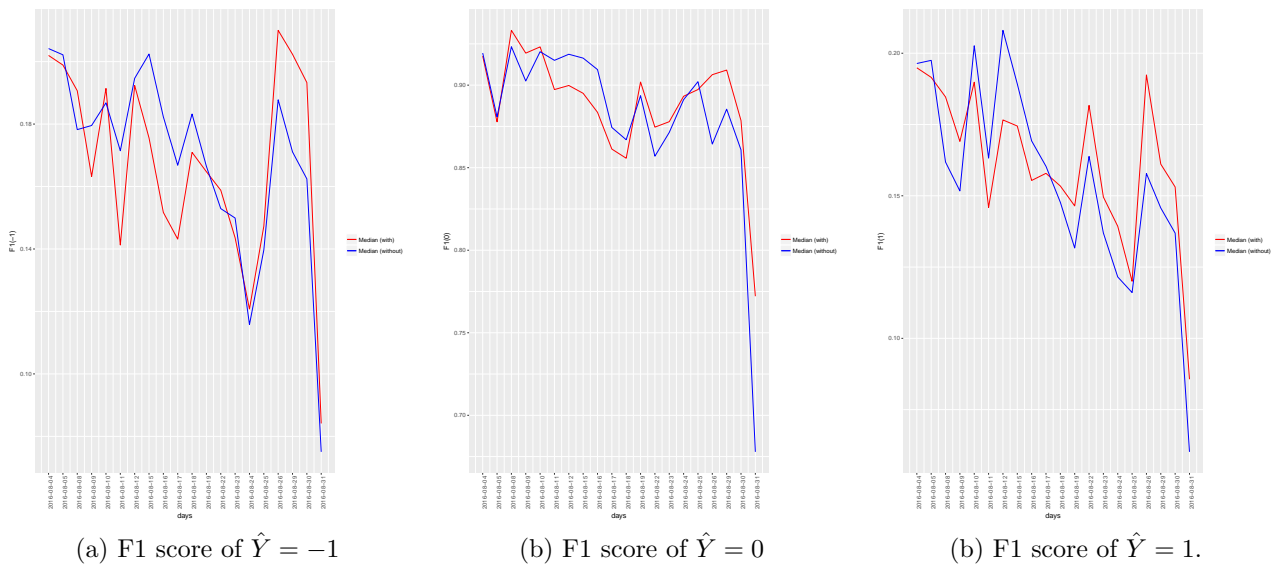

(a) F1 score of $\hat{Y} = -1$    (b) F1 score of $\hat{Y} = 0$    (b) F1 score of $\hat{Y} = 1$.

Figure 8: *The F1 scores over the calendar month, with (red) and without (blue) daily retraining of the RNN, are shown for (left) downward, (middle) neutral, or (right) upward next price movement prediction.*

Figure 9 compares the Receiver Operator Characteristic (ROC) curves of a binary RNN classifier over varying prediction horizons. The plot is constructed by varying the probability threshold (a.k.a. cut-points) for positive classification $Y = 1$ over the interval $[0.5, 1)$ and estimating the true positive and true negative rate of each model. The dashed line shows the performance of a white-noise classifier. We observe that the performance of the RNN decays as we increase the prediction horizon from the next book update event to 1s in the future. Latency between trading systems and the exchange is dependent on several factors including the execution platform, the distance of the co-located server to the exchange and even the amount of incoming traffic to the CME matching engines. It is commonplace however for this latency to lie in the range of 100 micro-seconds to 1ms. Our estimate, based on preliminary investigation, is that a c/c++ implementation of our RNN can predict in around 100 micro-seconds on a modern CPU with compiler optimizations but without low level programming techniques for code optimization.

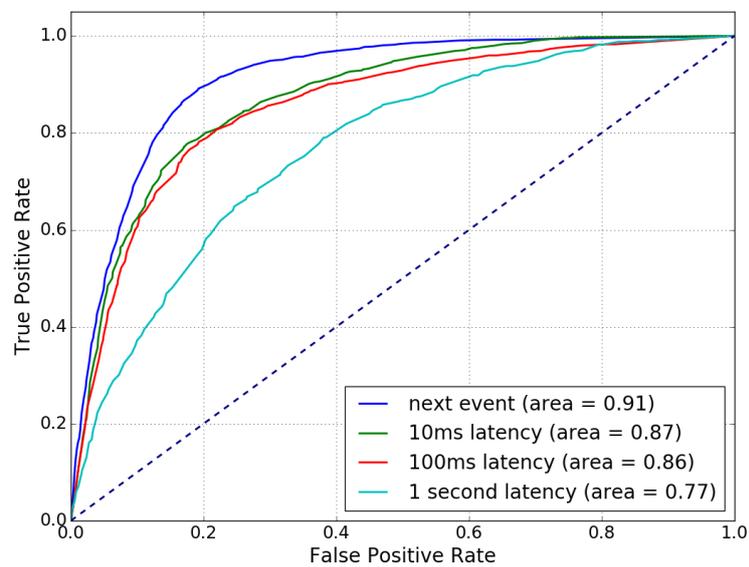

Figure 9: *The Receiver Operator Characteristic (ROC) curves of a binary RNN classifier over varying prediction horizons. In practice, the prediction horizon should be chosen to adequately account for latency between the trade execution platform and the exchange.*

# 6   Conclusion

This paper introduces a high frequency trade execution model to evaluate the economic impact of supervised learners. We present a trade information matrix to attribute the expected profit and loss of tick level predictive classifiers. This attribution considers execution constraints such as fill probabilities and position dependent trade rules. Our findings suggest that our approach can be a valuable and computationally tractable addition to standard market simulation based backtesting. In particular, it allows strategies to be compared given the confusion matrix of the classifier and could even be used to tailor a strategy to the accuracy characteristics of the classifier. Some of these characteristics have been presented for ESU6 over the month of August 2016.

For supervised learning and our trade information matrix to be of practical use in market making, a number of questions remain as a corollary of this research. Some of these questions relate to the supervised learner itself. For example, how can the prediction label $Y_t$ account for the impact of a trade on the state of the book when the decision to trade is itself based on a prediction $\hat{Y}_t$ of that label? Is reinforcement learning the best approach for addressing this issue, as described for market taking in Kearns and Nevmyvaka (2013)? How much latency does a RNN model add to the quote time if implemented efficiently? Then there are questions surrounding the trade information matrix; How frequently should the fill probability estimates be updated? Should fill probabilities be conditioned on time of day and, perhaps, other periodic events such as earnings announcements?

Our preliminary findings suggest that there is merit in applying non-linear classifiers, such as RNNs, to Level 2 data. Instead of asking the question 'can supervised learning be used for high frequency market making?', we believe that this paper advances this discussion to a question of how much error is tolerable when supervised learning is used to avoid adverse selection.